# Complementarity paradox solved: surprising consequences

**E V Flores**
**J M De Tata**
Department of Physics & Astronomy
Rowan University
Glassboro, NJ 08028
Email: flores@rowan.edu

**Abstract.** Afshar et al. claim that their experiment shows a violation of the complementarity inequality. In this work, we study their claim using a modified Mach-Zehnder setup that represents a simpler version of the Afshar experiment. We find that our results are consistent with Afshar et al. experimental findings. However, we show that within standard quantum mechanics the results of the Afshar experiment do not lead to a violation of the complementarity inequality. We show that their claim originates from a particular technique they use to analyze their results. In their analysis, they assume a classical concept, that particles have a definite trajectory before detection, thus, they obtain which-way information by particle detection plus path extrapolation by applying momentum conservation. This analysis technique is standard in experimental particle physics. Important discoveries such as the detection of vector bosons have been made through the application of this technique. We note that particle detection plus path extrapolation is a suitable technique within de Broglie-Bohm theory of quantum mechanics.

## 1. Introduction

The which-way information parameter $K$ measures the distinguishability among possible paths available to a particle. If the particle is equally likely to take any of the available paths then $K$ is 0. If a particular path is much more likely than any other path then $K$ is close to 1. Determining the path of a particle is a particle-like property; thus, when $K$ is 1 we claim that we know the path of the particle. On the other hand the visibility parameter $V$ measures the degree of wave interference; thus, when the visibility is 1 we have observed full destructive interference, a unique wave phenomenon. However, it is possible to obtain partial which-way information and visibility as long as the results fall within the complementarity inequality [1,2],

$$K^2 + V^2 \le 1\,. \tag{1}$$

The Afshar experiment [3,4] is a type of two-slit experiment with results that show an apparent violation of the complementarity inequality in Eq. (1). The radical claim of Afshar et al. implies observation of physical reality in the classical sense for both particle-like $(K)$ and wave-like $(V)$ properties of photons in the same experimental setup. Their claim of a violation of the inequality in Eq. (1) has produced strong criticism. Researchers seeking a resolution of the paradoxical findings have resorted to discredit the validity of Afshar et al. analysis [5-8].

In this paper we attempt to clarify the confusion caused by the claims of Afshar et al. We analyze the modified Mach-Zehnder setup of Fig. 1 due to its relative simplicity and transparency for calculation and analysis purposes. The modified Mach-Zehnder of Fig. 1 is an optical equivalent of the Afshar experiment [9]. Our modified Mach-Zehnder setup consists of a laser beam that impinges on a 50:50 beam splitter and produces two spatially separated coherent beams of equal intensity. Ideally there should only be one photon in the entire setup at a time. The beams overlap at some distance and then separate again. Detector 1 is positioned in front of mirror 1, and detector 2 is positioned in front of mirror 2. Where the beams overlap they interfere and form a pattern of bright and dark fringes. At the center of the dark fringes we may place a wire-grid made of thin absorbent wires. The purpose of the wire-grid is to verify the presence of interference fringes with minimal disturbance to the beams.

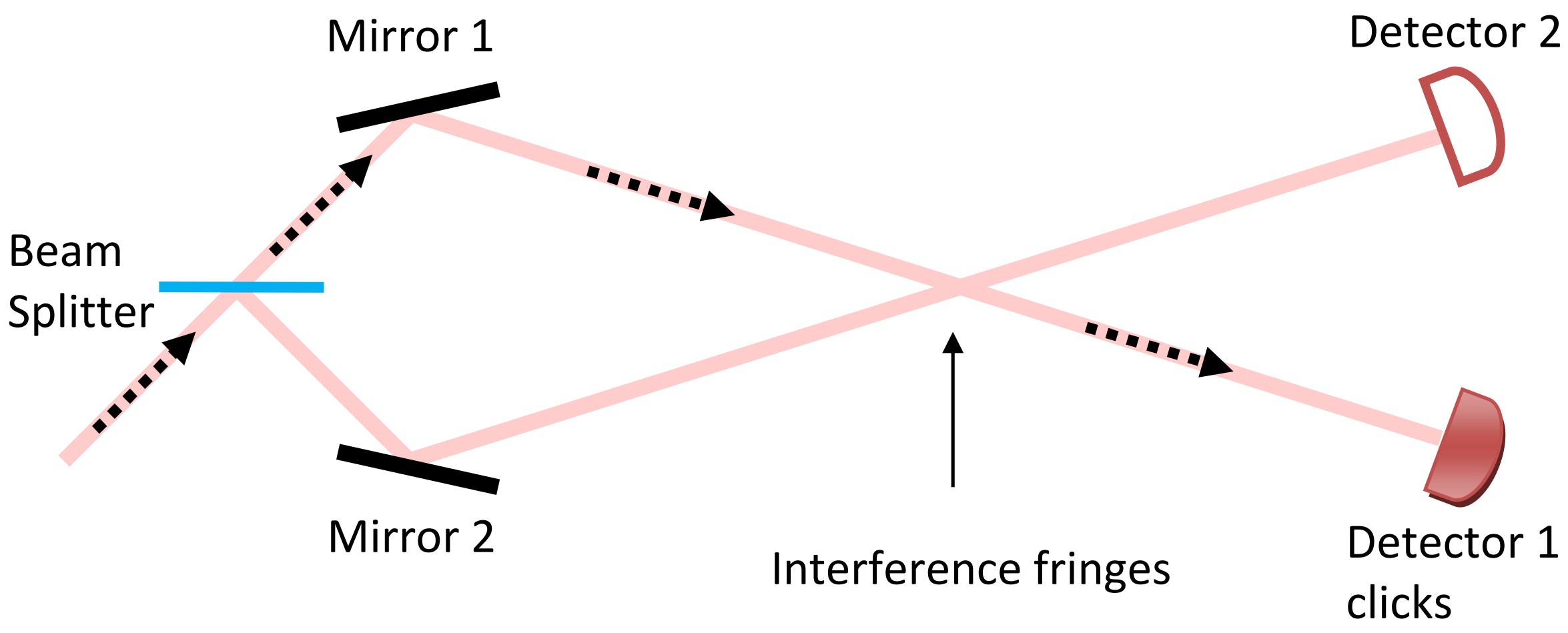

Fig. 1. This modified Mach-Zehnder setup is an optical equivalent of the Afshar experiment. A laser beam impinges on a 50:50 beam splitter and produces two spatially separated coherent beams of equal intensity. Where the beams overlap they interfere and form a pattern of bright and dark fringes. At the center of dark fringes thin wires

may be placed. With a single photon in the entire setup only one detector clicks. If detector 1 clicks we may trace the path of the photon with arrows as shown in picture.

When the wire-grid is removed the two possible paths for the photon are free of obstacles. Conservation of linear momentum allows us to trace uniquely the path of the photon. Thus, if detector 1 clicks we assume that the photon came from mirror 1. Similarly, if detector 2 clicks we assume that the photon came from mirror 2. This procedure, path extrapolation using momentum conservation, is standard in the analysis of particle physics experiments [10].

A prominent example of criticism of the work of Afshar et al. is found in the work of Jacques et al. [5]. Jacques et al. claim that Afshar et al. did not realize that photons which are identically prepared break into disjoint subsets according to where they land [5]. According to Jacques et al., disjoint subsets of photons should not be mixed in a calculation of complementarity aspects such as visibility $(V)$ and which-way information $(K)$. A similar observation was proposed before by Steuernagel [6]. Jacques et al. and also Steuernagel question the validity of Afshar et al. technique to measure the visibility. Flores replied to Steuernagel's paper pointing out that photons which are identically prepared have identical amplitudes for each possible outcome; where a photon actually lands, wires or detectors, is a probabilistic event [11]. Equally prepared photons are in a pure state and do not fall into disjoint subsets according to where they land. In conclusion, all equally prepared photons belongs to a single set that contributes to a measurement of the which-way information $(K)$ and visibility $(V)$.

Kastner has argued in favor of the visibility measurement technique used in the Afshar experiment [8]. However, Kastner argued against the claim of Afshar et al. that they have evidence of a violation of the complementarity inequality. Kastner pointed out that when the two possible photon paths of Fig. 1 are open we cannot assume that the photon only took one path. The assumption that the photon took a unique path is not quantum mechanical but semiclassical. It is not surprising that the application of this semiclassical analysis lead Afshar et al. to their conclusion that their experimental results imply a violation of the complementarity inequality, Eq. (1).

At the present time there is not agreement among researchers on the resolution of the Afshar experiment paradox. Several ideas have been proposed but they have been refuted or not fully embraced by the majority [5-8,11,13]. In our view Kastner has come closer to a satisfactory solution of the paradox. Our present work is based on Kastner's observation of a problem with the way Afshar et al. obtain which-way information. An experiment with single photons carried out by Jacques et al. was expected to put to rest the controversy surrounding the Afshar experiment [5]. However, in their experiment Jacques et al. did not use the key ingredient of the Afshar experiment, namely, placing thin wires at the center of dark fringes to obtain the visibility [3]. As we mentioned before they do not believe that the visibility could be obtain in this way. Instead Jacques et al. obtain the visibility by scanning the grid through the diffraction pattern. Thus, even if their experiment and results are correct they do not resolve the Afshar experiment paradox.

Jacques et al. and also Steuernagel seem to favor the technique that consists of particle detection plus path extrapolation through momentum conservation to obtain which-way information [5,6]. However, they argue that in the Afshar experiment the application of this technique is limited by diffraction effects introduced by the wires. Thus, a measurement and calculation of wire diffraction is required. In section 2 we present a summary of the calculation of the Afshar experiment using the modified Mach-Zehnder setup of Fig. 1. In section 3 we show that within standard quantum mechanics the results of the Afshar experiment do not lead to a violation of the complementarity inequality. In section 4 we show how the semiclassical technique in Fig. 1 leads to a violation of the complementarity inequality. The semiclassical technique illustrated in Fig. 1 is standard in the analysis of experiments in particle physics and quantum optics experiments such as the delayed choice experiment. The application of this semiclassical analysis to some physics experiments is presented in section 5. In section 6 we consider some basic concepts of de Broglie-Bohm theory [12]. We point out how the results of the Afshar experiment seem to support the realistic view of wave and particle in de Broglie-Bohm theory. In section 7 we summarize our findings.

## 2. Hilights of calculation

In the modified Mach-Zehnder setup of Fig. 1 thin absorbent wires are introduced at the center of the dark fringes where the two beams cross [13]. We use thin wires to measure the visibility of the interference pattern with minimal disturbance to the beams [3]. A calculation of the effect of the wire-grid on the beams has been done before in detail [14]. However, the calculation presented here is an improvement of previous work. In the previous calculation the wire thickness is a constant; in the new calculation the wire thickness is a free parameter. In the new calculation the beam cross section has been changed from circular to square to allow the use of analytical results rather than numerical approximations for diffraction. Overall, the new results are consistent with the old except that the new calculation fixes a mistake which does not change any of our conclusions.

Our calculation is based on Fraunhofer diffraction [15]. The key intermediate step in the calculation is diffraction of light due to the complementary screen to the wire-grid, a slit-grid [15]. The slit-grid is easier to study. Light diffracted by the slit-grid turns out to be similar to light diffracted by the wire-grid except at the detector region where the electric field of the original beam needs to be included in the calculations. This is a consequence of Babinet's principle which is an expression of the linearity of the fields [15]. Another consequence of Babinet's principle is that the total number of photons absorbed by the wire-grid is similar to the number of photons diffracted by it.

In our calculation we use light with $\lambda = 638$ nm. The thickness $b$ of six identical wires is a free parameter; the center to center wire separation is $d = 319$ μm, and the beam has a square cross section of side $2.55$ mm. Two coherent laser beams propagate symmetrically on a plane and cross each other at an angle of $0.002$ rad. For the particular case when $b = 32$ μm we can compare the calculation with the experimental results of Afshar et al. [4,9].

We first calculate the effect of the wire-gird on a single beam approximated by a monochromatic wave with finite cross-section [22]. This calculation is based on the standard slit-grid diffraction [15]. For the particular case, $b = 32$ μm, we find that the decrease in photon count at the detector in front of the beam is significant $(14.38\%)$. The wires absorb or reflect a number of photons $(7.53\%)$. We also find that some of the photons $(0.66\%)$ are scattered to the wrong detector. These values agree with experimental values [4,9].

When the two beams are on, in the region where the beams intersect we expect the formation of an interference pattern with bright and dark fringes. This could be easily verified using an opaque screen. We now place thin wires at the center of the dark fringes and remove the opaque screen but leave the wires in place. We find that significant diffraction occurs only on the plane that contains the beams [22]. Thus, our purpose is to obtain the intensity of diffracted light as a function of the angle $\theta$. The angle $\theta$ is on the plane that contains the beams. The midpoint between the two detectors in Fig. 1 corresponds to $\theta = 0$. We find that the intensity is given by [14]

$$I(\theta) = \frac{\Lambda}{(\kappa \sin\theta)^4} \left\{ b\kappa \sin\theta \cos\left(\frac{b\kappa\sin\theta}{2}\right) - 2\sin\left(\frac{b\kappa\sin\theta}{2}\right) \right\}^2 \times$$

$$\left\{ \sin\left(\frac{d\kappa\sin\theta}{2}\right) - \sin\left(\frac{3d\kappa\sin\theta}{2}\right) + \sin\left(\frac{5d\kappa\sin\theta}{2}\right) \right\}^2, \tag{2}$$

where $\Lambda$ is a constant and $\kappa = \frac{2\pi}{\lambda}$; $b$ and $d$ were defined before. We note that this result has been confirmed experimentally [9]. In Fig. 2 we plot the intensity in Eq. (2) for the case when $b = 32$ μm and $d = 319$ μ m. The location of the first peak, at either side of the interference pattern, coincides with the location of the detectors. We note that the area under the first peak is only $0.075\%$ of the total area, thus, most light is diffracted away from the detectors.

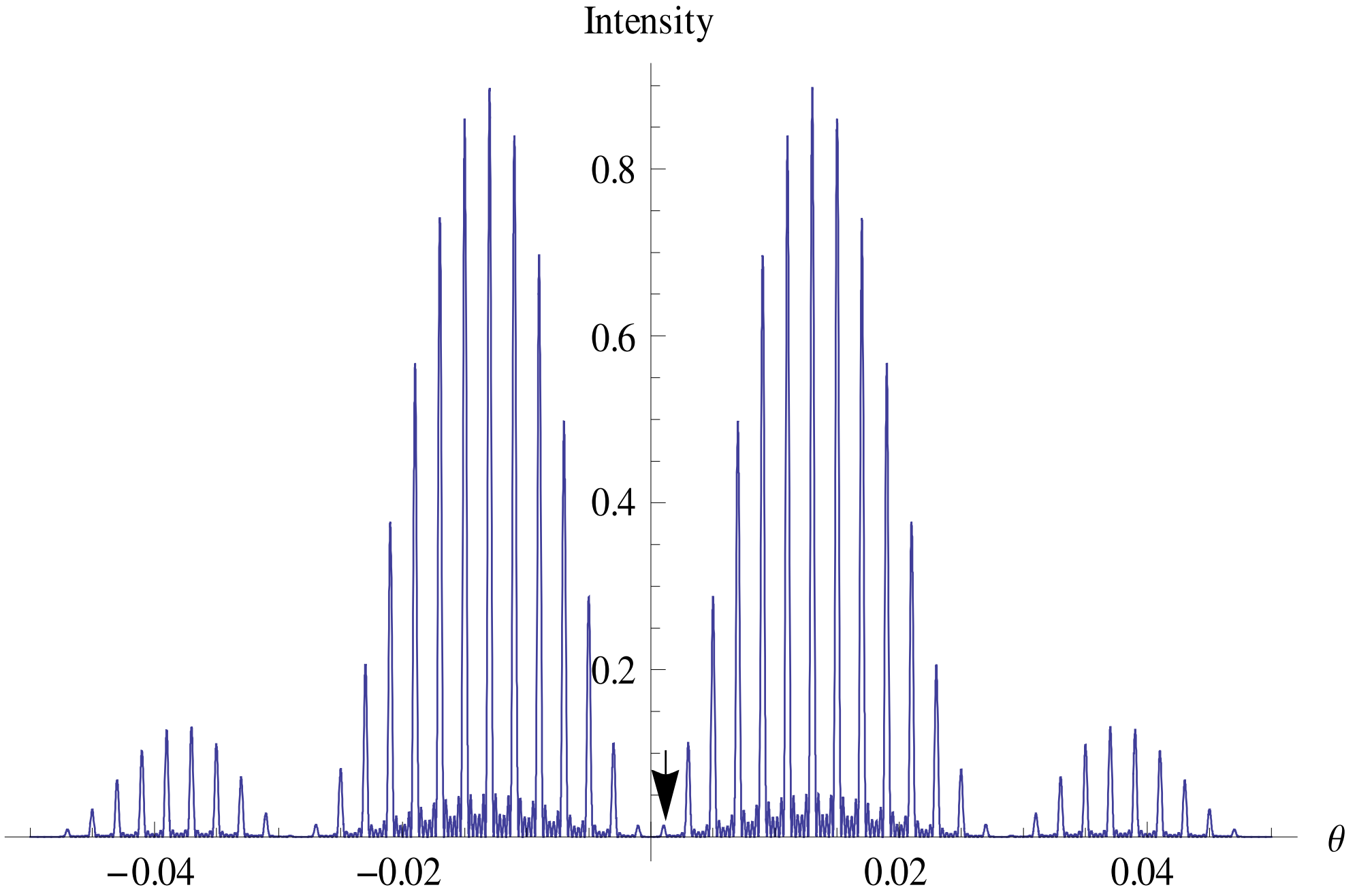

FIG. 2. Diffraction pattern for two coherent beams that interfere and diffract from a slit-grid located at the center of dark fringes. Notice that detectors 1 and 2 are symmetrically located at the first peak from the center, $\theta = \pm 0.001$ rad. The arrow shows the location of one of the detectors. This diffraction pattern has been observed experimentally.

Using $I(\theta)$ in Eq. (2) and conservation of energy we get results that can be compared with experiment. The percentage of light diffracted away from both detectors is small $(0.1238\%)$. The wires absorb or reflect a portion of the beam $(0.1240\%)$. Thus, we find that the presence of wires hardly decreased the photon count $(0.2478\%)$ at either detector as compared to the case without the wire-grid. These results are consistent with the previous calculation and with experimental values [14,4]. Finally, we find that the percentage of diffracted light that falls on either detector is very small $(0.0002\%)$. We discovered that due to a trivial error in previous calculation this number $(0.0002\%)$ is 5 times smaller than the previously reported number $(0.001\%)$ [14]. We note that the percentage of diffracted light that falls on either detector cannot be measured directly. However, the result of the calculation $(0.0002\%)$ helps us to appreciate that only two photons in a million are deflected by the wire-grid towards the detectors.

## 3. Quantum analysis

The quantum mechanical predictions for the modified Mach-Zehnder setup may be broken into two cases: *with* and *without* a wire-grid. When no wire-grid is present the two beams in Fig. 1 cross and then separate unchanged. Since the two beams are identical and remain unperturbed then the probability that the particle is in either beam is ½. In this case we cannot distinguish which-way a particle would go, thus, the which-way information parameter is zero [2,16], $K = 0$. We note that when the visibility parameter $V$ is not measured the standard practice in quantum mechanics is to assign its lowest value of zero. Therefore, the quantum mechanical prediction for this case is

$$K^2 + V^2 = 0 . \tag{3}$$

When the wire-grid is in place we find that the wires with $b = 32$ μm hardly decrease the photon count $(0.2478\%)$ at either detector as compared to the case without the wire-grid. As mentioned before, the wires would produce a significant drop in photon count $(14.38\%)$ only if the net beam was uniform, not showing interference fringes. Thus, we conclude that the small decrease in photon count $(0.2478\%)$ when the wire-grid is in place and the two beams are on is evidence that at the intersection, where the beams meet, the resultant beam is not uniform but it is the result of wave interference with high visibility. Unfortunately, we cannot measure the visibility directly but we can place a lower limit consistent with experiment [4,14]. To calculate the lower limit for the visibility we need to know, $N$, the number of photons that go from a mirror towards its corresponding detector and the number of photons stopped by the wires, $xN$. These numbers are easily obtained from photon count measurements at the detector region without and with the wire-grid in place.

For a numerical example of the calculation let us assume that $N = 10^6$, then, the number of photons stopped by the wires of thickness $b = 32$ μm would be $1{,}240$ $(0.1240\%)$. The $1{,}240$ photons that are stopped by the wires are part of the minimum intensity region while the remaining $998{,}760$ photons that go through are part of the higher intensity region. We start by assuming ignorance about the shape of the interference pattern. We consider the standard formula for the visibility [15]

$$V = \frac{I_{\max} - I_{\min}}{I_{\max} + I_{\min}}, \tag{4}$$

where $I_{\max}$ and $I_{\min}$ are the maximum and minimum intensities of the interference pattern. To minimize the visibility $I_{\max}$ needs to be as small as possible and $I_{\min}$ as large as possible. To maximize $I_{\min}$ the darker regions must have the geometrical shape of thin rectangular boxes each with a base equal to the thickness of the wire, $0.032$ mm, times its corresponding length $2.55$ mm. $I_{\min}$ is proportional to $1{,}240$ divided by the area of the base of the six thin boxes, $0.4896$ mm$^2$, $I_{\min} \propto 2{,}533$. Similarly, $I_{\max}$ is minimized by distributing uniformly the photons that miss the wires $(998{,}760)$ on an area of $6.0129$ mm$^2$ which is the beam cross section minus the area covered by the wires, thus, we get $I_{\max} \propto 166{,}103$. Therefore, the interference pattern with the lowest visibility for our setup is a type of periodic square function. Using Eq. (4) we get a lower limit for the visibility, $V \geq 0.9699$.

We may apply the same technique to obtain a theoretical lower limit for the visibility of the interference pattern using wires of arbitrary thickness $b$. The total number of photons that come from a given mirror towards its corresponding detector is $N$. The number of photons that are stopped by the wires, $xN$. $A$ is the cross section of the beam and $yA$ is the portion of the area cover by the wire-grid. Therefore, $I_{\min}$ is proportional to $xN/(yA)$ and $I_{\max}$ is proportional to $(1-x)N/[(1-y)A]$. Putting these expressions in Eq. (4) we get

$$V \geq \frac{(1-x)/(1-y) - x/y}{(1-x)/(1-y) + x/y} \tag{5}$$

The fractions $x$ and $y$ may be easily obtained from experiment. However, we can also obtain a theoretical value in terms of the wire thickness $b$. Presuming that the wires are at the minima of an interference pattern we calculate $x$ by integrating the intensity of the field over the area covered by the wires and divide this number by the integral of the intensity over the total beam cross-section. We find $y$ by taking the ratio of the cross section of wire-grid over the cross section of the beam. In Fig. 3 we plot the lower limit for the visibility squared as a function of the wire thickness $b$. We note that as the wire thickness decreases the visibility squared approaches its maximum value of 1. However, our technique is limited to wires of finite thickness since at the limit of zero thickness there is no measurement.

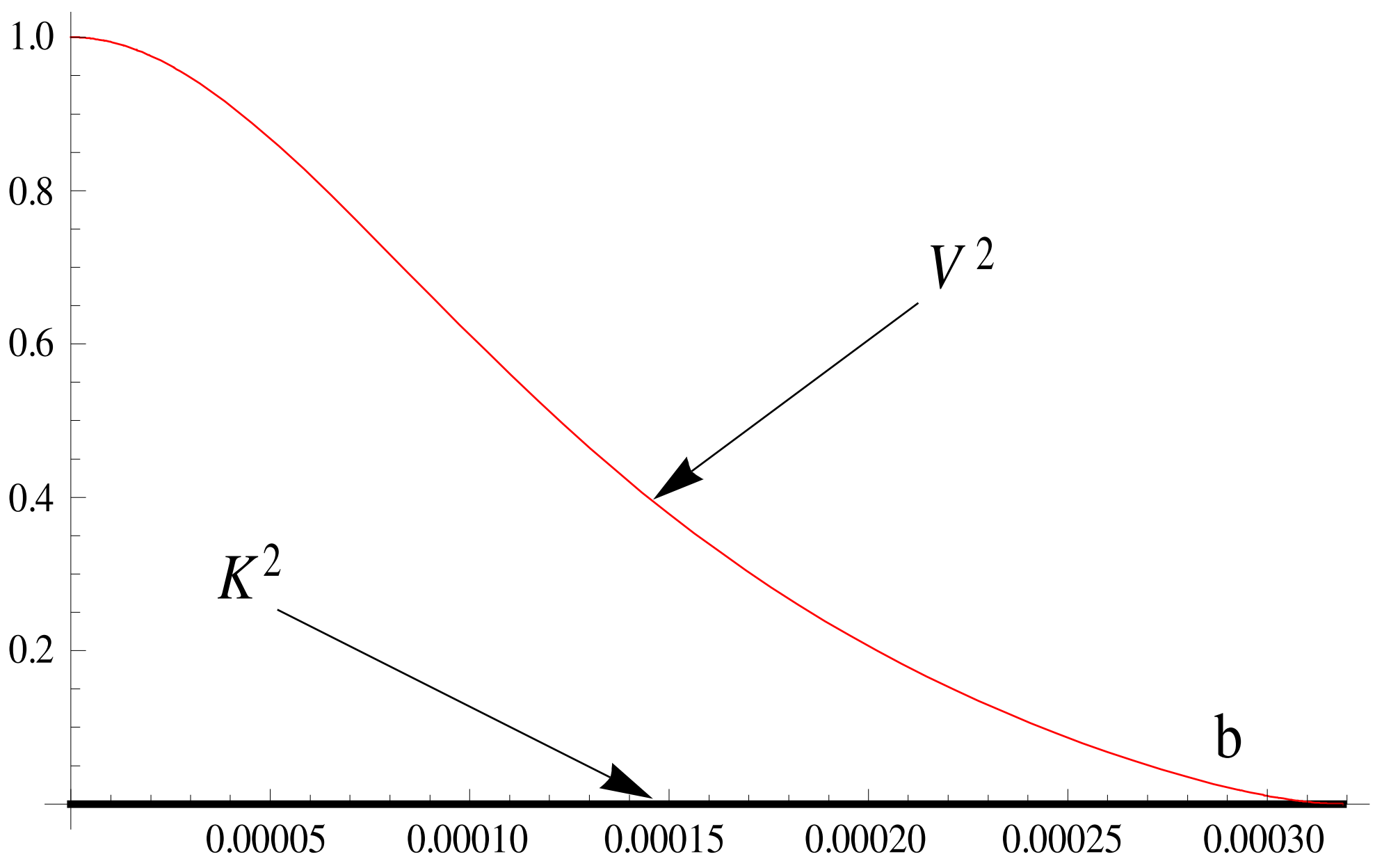

FIG. 3. The lower limit for the visibility squared $V^2$ and the which-way information squared $K^2$ are plotted as a function of the wire thickness $b$. The sum of the two is less than one in accordance with the complementarity relation.

To conclude the quantum mechanical prediction for the case *with* the wire-grid in place we estimate the which-way information. We notice that the wire-grid is symmetric and it is also symmetrically placed at the beams intersection. Thus, the wire-grid has an identical effect on both beams. The probability of finding a photon in either beam is the same. Therefore, there is no way to distinguish which-way a photon would go, thus, the which-way information parameter is zero, $K = 0$. Adding our results for the visibility and the which-way information we find that our results are in agreement with the complementarity inequality, Eq. (1), for thin wires of arbitrary thickness.

## 4. Semiclassical analysis

Sometimes we might need to know what took place before particle detection; we might do this by extrapolating to regions where no detection was possible. In some experiments we often can only measure portions of the path of outgoing particles. With this information, applying conservation laws, we extrapolate the path of the particle and determine its origin. We apply this technique to obtain path information in our modified Mach-Zehnder setup *without* the wire-grid, Fig. 1. In this case the two beams are undisturbed all the way from beam splitter to the corresponding detector. With a single particle present in the setup only one detector clicks. A click of the detector automatically determines the location and momentum of the particle; by applying momentum conservation, we obtain the path consistent with the detected value. To distinguish this result from the quantum mechanical which-way information parameter $K$ we introduce a new parameter, $K'$. We define the parameter $K'$ as the classical which-way information or distinguishability of the path of the particle. In our example we apply momentum conservation and obtain full classical which-way information, $K'=1$. This result is very different from the quantum mechanical prediction, $K=0$, for this case. However, there is no paradox here as the two parameters refer to different situations. The standard parameter $K$ represents either a prediction or a direct measurement of which-way the photon would go. The new parameter $K'$ represents inferred path information through momentum conservation.

*With* the wire-grid in place and the two beams on we already obtained a lower limit for the visibility. For the particular case of wire thickness $b=32$ $\mu\text{m}$ we got $V \geq 0.9699$. In this case, a click of a given detector provides only partial which-way information for the photon due to the presence of the wire-grid in the path of the photon. However, we are certain about the path of the photon from the wire-grid to the detector that clicked. To go further, beyond the wire-grid, we need to estimate the effect of the wire-grid on the photon momentum. Going back to our calculation we consider $10^6$ photons that come from a mirror towards its corresponding detector one at a time. On the average the detector will collect $997{,}522$ of them. The losses can be easily accounted for. The wires stop $1{,}240$ photons. The total number of diffracted photons away from the detector is $1{,}238$. The sum of these two numbers accounts for the $2{,}478$ decrease in photon count at the detector. The calculation also shows that on the average only $2$ photons out of $10^6$ are diffracted towards the detector. We notice that diffracted light does not have which-way information as its origin could be either mirror. Fortunately, most of the diffracted photons $(1{,}238)$ do not reach the detectors. Since only two out of a million photons are diffracted towards the detector, the wire-grid has an insignificant effect on the momentum of the photon that reaches the detector. Thus, the classical which-way information parameter $K'$ is close to its maximum value of 1.

To obtain a lower limit for $K'$ we need to find how many of the initial $N$ photons that come from a given mirror towards its corresponding detector have not been deflected or absorbed by the wire-grid. The number of photons stopped by the wires is $xN$; thus, we subtract $xN$ from $N$. Due to Babinet's principle [15] the total number of diffracted photon is also $xN$; these photons have zero which-way information, thus, we subtract them form $N$. The remaining photons, $N-2xN$, are not deflected.

Therefore, a lower limit for the classical which-way information is

$$K' \geq \frac{N - 2xN}{N} = 1 - 2x \,. \tag{6}$$

At the limit of zero wire thickness no photons are stopped by the wires $(x = 0)$ and the classical which-way information in Eq. (6) is 1, as expected from previous consideration. At the other limit where the wires are thick enough to stop half of the photons we have that $x = 1/2$, then, according to Babinet's principle, the other half that goes through are all diffracted photons which implies that they have no classical which-way information. At this limit Eq. (6) gives a consistent result, $K' = 0$. In Fig. 4 we plot the classical which-way information squared, $K'^2$, as a function of the wire thickness.

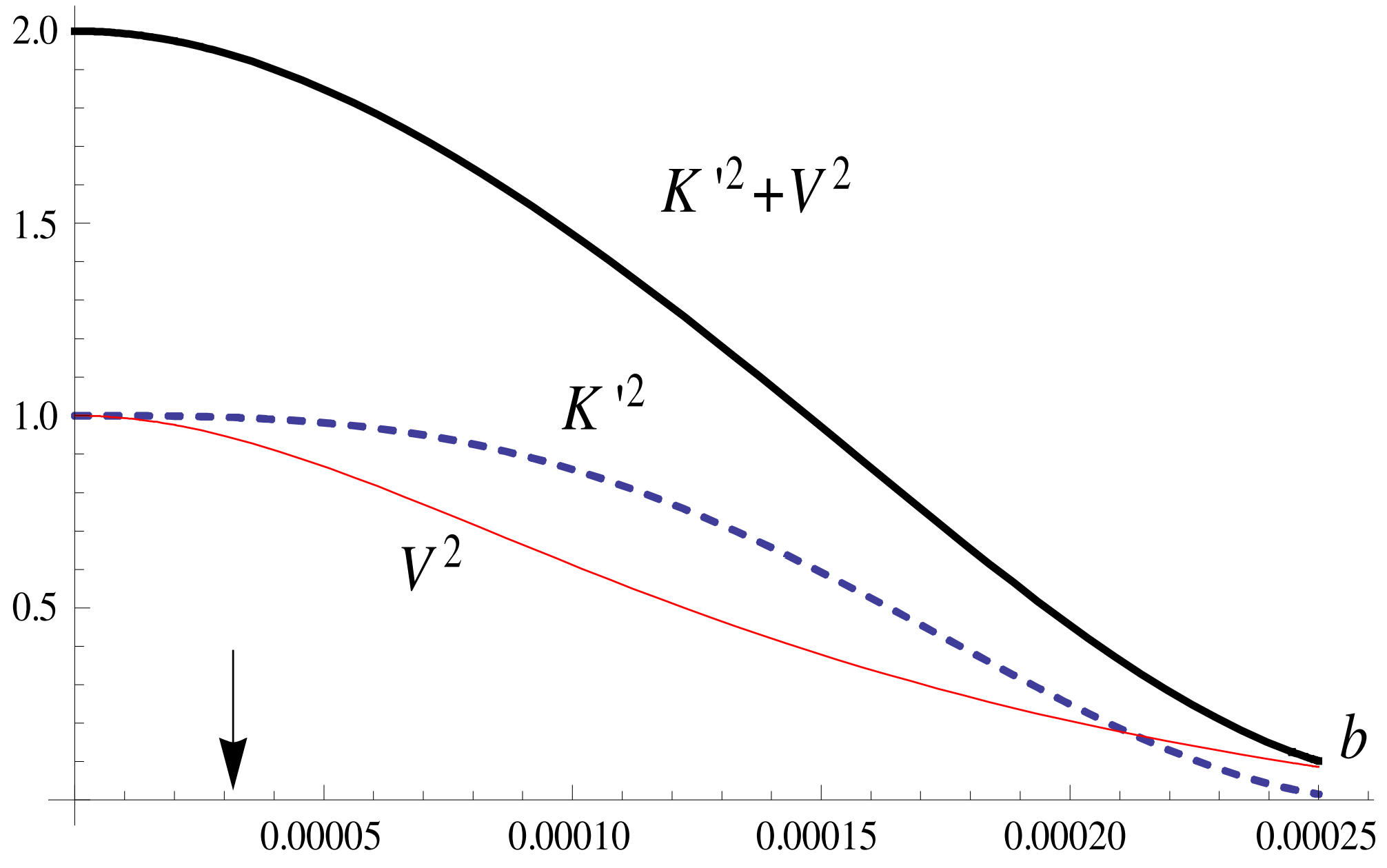

FIG. 4. Lower limit for the visibility squared, classical which-way information squared and their sum are plotted as a function of the wire thickness $b$. The arrow indicates the value of the wire thickness, $b = 32$ μm, for which there is experimental evidence in the optical equivalent experimental setup.

The results in Fig. 4 for the modified Mach-Zehnder setup *with* the wire-grid in place suggest the existence of an inequality

$$K'^2 + V^2 < 2 \tag{7}$$

where $K'$ is the classical which-way information and $V$ is the standard visibility. Notice that in our inequality, Eq. (7), we do not use the symbol smaller than or equal to $(\leq)$ since in our experiment the which-way information $K'$ would be 1 only at the limit of zero wire thickness, however, at this limit we cannot measure the visibility so we set it to zero. We note that Kolar et al. [17] have proposed another complementarity inequality which they show is violated in their setup, while the standard complementarity inequality, Eq. (1), is not violated as in our case. However, an experiment to demonstrate their findings has not yet been performed.

## 5. The semiclassical technique in other experiments

The semiclassical approach used here to obtain the classical which-way information comes from an application the conservation laws. In our approach, the conservation laws are used to extrapolate the classical path of a particle to regions where measurements are not possible or desirable. This technique is standard in the analysis of particle physics experiments [10]. In high energy particle physics experiments it is sometimes impossible to directly detect a particle due to its short lifetime, as in the case of massive vector bosons, or due to its asymptotic properties such as in the case of quarks. In these and other cases we rely on information obtained from scattering of particles and decay products. The conservation laws allow us to extrapolate and obtain particle information such as charge, path, lifetime, mass, energy, and momentum of the otherwise undetected particles [10]. Obtaining information of where a particle was created and where it decayed through extrapolation of the conservation laws is a powerful technique that is based on classical concepts such a location or path.

Wheeler's delayed choice experiment [18] is another example where the application of the semiclassical technique is crucial. A version of the delayed choice experiment uses the Mach-Zehnder set up in Fig. 1 with the beams crossing at 90 degrees before reaching the end detectors. Where the beams cross we may place a 50:50 output beam splitter. The purpose of the beam splitter is to produce total destructive interference along beam that leads to detector 1 (or 2). This is confirmed by the fact that in this setup only detector 2 (or 1) clicks. Using Eq. (4) we see that the presence of total destructive interference implies full visibility, $V = 1$. However, in this case we cannot tell which path the photon took, $K' = 0$. If we choose not to include the output beam splitter in the set up then the path the photon took is easily revealed. A click of a detector locates the photon and momentum conservation reveals its path, thus, we have $K' = 1$. However, without the beam splitter we do not measure wave interference and we set $V = 0$. The choice to place or not to place the output beam splitter is made after a single photon has entered the interferometer.

We note that when we choose not to use the output beam splitter in Wheeler's delayed choice experiment the classical which-way information is maximum, $K' = 1$. However, the quantum mechanical which-way information for this case is zero, $K = 0$. The reason is that the two paths for the photon are identical all the way from the entrance beam splitter to the detectors. Since there is no way to distinguish which path the photon would take, the distinguishability of the ways or which-way information is zero [2]. Thus, the success of Wheeler's delayed choice experiment hinges on the application of the semiclassical technique. Finally, we note that Wheeler's delayed choice experiment has been experimentally realized [19]

**6. The de Broglie-Bohm theory**

The de Broglie-Bohm theory is a quantum theory describing the motion of point particles with definite trajectory [12]. The position changes according to the guiding equation. The guiding equation gives the velocity of the particle in terms of the gradient of the wavefunction. Schrödinger's equation completes the dynamical specification of this theory as it gives the evolution of the wavefunction. The evolution of the wavefunction does not depend on the position of the particles.

We note that the Afshar experiment has been done with light while de Broglie-Bohm theory has been fully developed to the non-relativistic level only. However, the Afshar experiment could be done with non-relativistic matter. The Afshar experiment only involves particle and wave aspects of matter and radiation and as far as we know the symmetry of matter and radiation in this respect has been established [20]. Thus, we expect that the results of the Afshar experiment with non-relativistic matter should be similar to the results of the Afshar experiment with light.

The de Broglie-Bohm theory reproduces all the results of the standard non-relativistic quantum mechanics without the paradoxes associated with the Copenhagen interpretation. For instance the particle-wave duality paradox is resolved in a simple way, namely, particle and wave both are real. Since the point particle is real it has a definite position and velocity at any time. According to D. Durr et al., "the role of the wave function in Bohmian mechanics is to *tell the matter how to move*" [12]. Thus, the wave function also has physical reality. Simultaneous physical reality for the wave and particle implies that the complementarity inequality in Eq. (1) could be violated in de Broglie-Bohm theory. However, since de Broglie-Bohm theory has similar predictions as standard quantum mechanics it is not easy to find an experiment to observe a violation of the complementarity inequality, Eq. (1). In experiments where complementarity could be associated with the uncertainty principle [21] or with quantum entanglements [2] we do not expect to see any discrepancy between standard quantum mechanics and de Broglie-Bohm theory.

The Afshar experiment turns out to be an experiment where complementarity is not associated with the uncertainty principle or with quantum entanglements [4]. Thus, standard quantum mechanics and de Broglie-Bohm theory could differ in their predictions for complementarity for the Afshar experiment. As discussed above, standard quantum mechanics predicts for the Afshar experiment with a wire-grid a high visibility, $V \approx 1$, but zero which-way information, $K = 0$, so that there is no violation of the complementarity inequality, Eq. (1). On the other hand, de Broglie-Bohm theory predicts a similar visibility, $V \approx 1$, but high which-way information, $K' \approx 1$, so that the complementarity inequality, Eq. (1), is violated. In de Broglie-Bohm theory the which-way information is high due to the fact that the particle has a continuous trajectory. If in the setup of Fig. 1 detector 1 clicks then the particle was in the neighborhood moments before. Using this logic or the guiding equation we could construct the path before detection which in this case is going to be similar to the one predicted using the semiclassical measurement technique, particle detection plus path extrapolation applying momentum conservation.

**7. Summary and conclusions**

The quantum mechanical analysis of the which-way information for the modified Mach-Zehnder setup of Fig. 1 *without* the wire-grid appears to be in contradiction with the path information obtained from momentum conservation. The probability that the photon would take one of the two identical paths is ½, which results in zero which-way information, $K = 0$ [2]. On the other hand, a click of a detector together with an application of momentum conservation gives us full classical which-way information, $K' = 1$. To clarify the fact that there is no contradiction we introduce the classical which-way information parameter $K'$ for cases when we get path information by applying conservation laws and assuming the classical concept of trajectory.

Our analysis of the modified Mach-Zehnder setup of Fig. 1 *with* the wire-grid in place shows that it is possible to form an additional complementarity inequality, Eq. (7), that contains $K'$ rather than $K$. Thus, when the wire thickness is $b = 32$ $\mu\mathrm{m}$ we have two results which do not contradict each other: $0.941 \le K^2 + V^2 \le 1$, and $1.932 \le K'^2 + V^2 < 2$. The first inequality is the standard quantum mechanical result that shows that two identical coherent beams could lead to an interference pattern with high visibility, $V^2 \ge 0.941$. However, for two identical beams the distinguishability of the ways or which-way information is zero, $K = 0$. The second inequality also shows that two identical beams could lead to an interference pattern with high visibility. However, photon detection plus momentum conservation allows us to infer the path of particle accurately.

The impossibility of violating the complementarity inequality, Eq. (1), within the standard version of quantum mechanics is well established [2]. Thus, the complementarity violation observed in the Afshar experiment, $K'^2 + V^2 \approx 2$, may be seen as confirmation that the technique that consists on a click of a detector plus path extrapolation applying conservation of momentum is semiclassical, thus, it is not consistent with standard quantum mechanics. If this is the case then the success of this semiclassical measuring technique in fields like experimental particle physics [10] and quantum optics [19] requires explanation. We note that this semiclassical measuring technique is compatible with de Broglie-Bohm theory, thus, this measuring technique should not introduce paradoxes or problems to this theory. Since de Broglie-Bohm theory is a complete and consistent quantum theory with an output equivalent to the standard quantum theory we do not expect that the semiclassical measuring technique would introduce paradoxes or problems to the standard quantum theory either. Finally, we note that the results of the Afshar experiment seem to highlight the need for a theory along the line of de Broglie-Bohm theory. There are potential benefits for research in theories like de Broglie-Bohm theory. For instance, the requirement that particle and wave both have some sort of physical reality could encourage new areas of research.